\documentclass[
twocolumn,
prl,
preprintnumbers,superscriptaddress,showpacs
]{revtex4-1}

\usepackage{amsmath} 
\usepackage{graphicx}
\usepackage{dcolumn}
\usepackage{bm}

\usepackage{color}
\definecolor{dgreen}{cmyk}{1.,0.,1.,0.2}        
\definecolor{orange}{cmyk}{0.,0.353,1.,0.}    

\def \a {\alpha}

\def \e {\varepsilon}

\def \l {\lambda}

\def \si {\sigma}

\def \vk {{\bm{k}}}

\def \vA {\bm{A}}

\def \vu {\bm{u}}

\def \hk {\hat{\vk}}

\def \he {\hat{e}}

\def \le {\left}
\def \ri {\right}

\def \<{\langle}
\def \>{\rangle}
\def \+{\dagger}

\def \tphi {\tilde{\phi}}

\def \tN {\tilde{N}}

\def \tc {\tilde{c}}

\def \co {{\rm co}}

\def \tr {{\rm tr}}
\def \i {{\rm i}}
\def \e {{\rm e}}

\def \He {$^3$He }

\newcommand\sect[1]{\noindent \textit{\quad #1}---}

\begin{document}

\title{Berry Curvature and Spin-One Color Superconductivity}

\newcommand{\UIC}{Department of Physics, University of Illinois, Chicago, Illinois 60607, USA}
\newcommand{\LQT}{Laboratory for Quantum Theory at the Extremes, University of Illinois, Chicago, Illinois 60607, USA}

\author{Noriyuki Sogabe}
\affiliation{\UIC}
\affiliation{\LQT}

\author{Yi Yin}
\affiliation{
School of Science and Engineering,
The Chinese University of Hong Kong,
Shenzhen, 
Guangdong, 518172, China
}
%
\noaffiliation

\date{\today}%

\begin{abstract}

We explore the interplay between Berry curvature and topological properties in single-flavor color superconductors, where quarks form spin-one Cooper pairs. By deriving a new relation, we connect the topological nodal structure of the gap function in momentum space to the (nonabelian) Berry flux associated with paired quarks. This generalizes the early work by Li and Haldane~[Phys. Rev. Lett. {\bf 120}, 067003 (2018)] to systems with additional internal quantum numbers, such as color. In the ultrarelativistic limit, we uncover rich topological structures driven by the interplay of spin, chirality, and color. Specifically, we identify chirality-induced topological nodes in the transverse (opposite chirality pairing) polar and {\it A} phases. In contrast, the color-spin-locking phase lacks these nodes due to a nontrivial color Berry flux, which in turn induces gapless excitations with total Berry monopole charges of $\pm 3/2$—differing from conventional Weyl fermions. Our findings can be potentially extended to other fermionic systems carrying additional internal degrees of freedom. 

\end{abstract}
\maketitle

In many-body systems involving chiral (Weyl) fermions, the Berry curvature associated with those particles leads to various interesting transport phenomena. One prominent example is the chiral magnetic effect (CME), where a current is generated along an applied magnetic field~\cite{PhysRevD.78.074033,PhysRevD.22.3080}. Recent theoretical advances revealed deep connections between the CME, Berry curvature, and quantum anomaly~\cite{Son:2012wh,Stephanov:2012ki,Jia_2016}. The consequence of the CME has been observed in Weyl and Dirac semimetals \cite{Li:2014bha,arnold2016negative,doi:10.1126/science.aac6089,PhysRevX.5.031023} and is under study in the quark-gluon plasma \cite{Kharzeev:2024zzm}. Additionally, Berry curvature induces the spin Hall effect~\cite{2003Science,Son:2012wh} and an analogous effect in the QCD plasma ~\cite{Liu:2020dxg,Fu:2022myl}, along with other novel spin-related phenomena~\cite{Liu:2021uhn,Fu:2021pok,Becattini:2021suc,Becattini:2021iol} actively studied through heavy-ion collision experiments~\cite{Becattini:2020ngo,Becattini2022}.

Much less attention has been devoted to the role of Berry curvature in cold and high baryon density QCD matter. At extreme densities, the ground state is a color-flavor-locking (CFL) phase, characterized by the Bardeen–Cooper–Schrieffer (BCS) pairing of three light quarks~\cite{Alford:1998mk}. At lower densities, factors such as a finite strange quark mass, electrical and color charge neutrality, $\beta$ equilibrium lead to Fermi-momentum mismatch and drive the system into various less symmetric phases of color superconductivity~\cite{Alford:1999pa,Alford:2000ze,Bedaque:2001je,Kaplan:2001qk,Alford:2003fq,Kundu:2001tt} (see Refs.~\cite{Alford:2007xm,Anglani:2013gfu} for reviews). Notably, single-flavor pairing of strange quarks or within all light quarks becomes plausible for a specific range of baryon densities~\cite{Schafer:2000tw,Alford:2002rz,Schmitt:2004et,Alford:2005yy}, with observational consequences explored in Refs.~\cite{Schmitt:2003xq,Wang:2009if,Wang:2010ydb}.

The attractive one-gluon exchange interaction, being color antisymmetric, naturally leads to a spin-one superconductor for single-flavor pairing. The interplay between spin and color gives rise to rich phases, including the polar, planar, {\it A}, and color-spin-locking (CSL) phases~\cite{Schafer:2000tw,Alford:2002rz,Schmitt:2004et,Alford:2005yy,Brauner:2008ma,Brauner:2009df}. Weak coupling single-flavor QCD calculations found that the (transverse) CSL phase is energetically favorable at asymptotically high density~\cite{Schafer:2000tw,Schmitt:2004et}. When single-flavor pairing is considered within three-flavor quark matter, the requirement of color neutrality may favor other patterns, such as that of the polar phase~\cite{Alford:2005yy}.

In the condensed matter literature~\cite{Murakami-PRL03,Li-Haldane-PRL18}, the ``pairing monopole charge" $\Delta q_{{\rm ch}}\equiv q_{{\rm ch}}-q'_{{\rm ch}}$ is introduced to characterize the Berry structure for the pairing state. For definiteness, we take $q_{{\rm ch}},q'_{{\rm ch}}=\pm 1/2$ to be the Berry monopole charges of the paired chiral fermions, though generalizing to a generic Abelian Berry monopole is straightforward. In Ref.~\cite{Li-Haldane-PRL18}, $\Delta q_{{\rm ch}}$ are connected to the sum of the topological number $g$ around the gapless points (nodes) of the energy gap at the Fermi surface (FS) (see details below),
\begin{align}
\label{eq:g-spin}
g=2\Delta q_{{\rm ch}} \,. 
\end{align}
This relation predicts topologically protected nodes when $\Delta q_{\rm ch}\neq 0$. The superconductivity with topological nodes has become a focal point of extensive research~\cite{Schnyder_2015,Sato_2017}.

In this Letter, we focus on spin-one superconductivity in the ultrarelativistic limit and explore the topological features of various pairing patterns. For light quarks, which are approximately regarded as chiral fermions, pairs can form with the same chirality (longitudinal), opposite chirality (transverse), or a mixture of both. While Eq.~(1) suggests that the transverse phase should exhibit topological nodes, a puzzle arises: some phases, like the polar and {\it A} phases, show point nodes, whereas others, such as the transverse CSL phase, remain fully gapped~\cite{Schafer:2000tw,Schmitt:2004et}. Why are the nodes absent in this phase despite $\Delta q_{{\rm ch}}\neq 0$?

We identify the missing element in the relation~\eqref{eq:g-spin}---the color structure. The fact that pairing fermions carry additional quantum numbers, such as color, significantly enriches the implications of nonzero $\Delta q_{{\rm ch}}$. By incorporating the color contribution, we obtain a generalized relation~\eqref{g-relation}. In the CSL phase, $\Delta q_{{\rm ch}}$ is canceled by the novel color contribution, yielding $g=0$ and hence the nodes can be absent. Moreover, we derive a remarkable relation~\eqref{q-zero} that implies an overlooked scenario in which the pairing monopole charges manifest themselves through topological gapless excitations.

\sect{Berry flux and the topological nodes}
We shall show Eq.~\eqref{g-relation}, which determines the relation between Berry flux of the pairing fermions and the sum of the topological number associated with the nodes, for a class of superconductors (including the spin-one color superconductor) described by the mean-field Hamiltonian in momentum space $\bm{k}$:
\begin{align}
\label{eq:gen-ham}
H= \int \frac{{\rm d}^3 \boldsymbol{k}}{(2\pi)^3}  \left(
\psi^\dagger \,,  \psi^{\prime\dagger}_{\rm c} 
\right)
\left(
\begin{array}{cc}
\mathcal {H}_0 & \Delta_{0}M \\
\Delta_{0} M^\dagger & 
\mathcal {H}_{0\rm c}^\prime
\end{array}
\right) \left(
\begin{array}{cc}
\psi \\
 \psi^{\prime}_{\rm c} 
\end{array}
\right)\,,
\end{align}
where $\psi$ and $\psi'$ are fermionic fields that may carry additional quantum numbers beyond spin, say color, and $\psi^{\prime}_{\rm c}(\vk)=\i\si_{2} \psi^{\prime\dagger}(-\vk)$ with Pauli matrices $\sigma_i\ (i=1,2,3)$. The real function $\Delta_0$ sets the overall magnitude of the gap and is independent of $\hat{\bm{k}}\equiv \bm{k}/k$ with $k\equiv|\bm{k}|$. We assume that there exist simultaneous eigenfunctions $\phi$ ($\phi^{\prime}$) of free Hamiltonian ${\cal H}_{0}$ and $M M^\dagger$ (${\cal H}_{0{\rm c}}^{\prime}$ and $M^{\dagger}M$), i.e.,
\begin{align}
\label{M-commute}
\le[M M^\dagger,{\cal H}_{0}\ri]=0\,,
\quad
\le[M^{\dagger}M, {\cal H}_{0{\rm c}}^{\prime} \ri]=0\, .
\end{align}
The Hermitian matrices $ M M^\dagger$ and $M^{\dagger}M$ share the same sets of nonnegative eigenvalues $\l$, which might depend on $\vk$. The pairing states are formed by single-particle states with the same nonzero eigenvalue $\lambda$, denoted as $\phi_{\lambda}$ and $\phi'_{\lambda}$ satisfying
\begin{align}
\label{eq:eigen_problem_MMd}
M M^\dagger\, \phi_{\l}&=\l\, \phi_{\l} \, ,
\quad
 M^{\dagger}M\, \phi'_{\l} =\l\phi' _{\l}\,.
\end{align}

If the degeneracy at given $\l$ is $N_{\lambda}$, we introduce a $N_{\lambda}$-dimension row vector $\Phi_{\lambda}=\{\phi_{\lambda,1},\ldots\}/\sqrt{N_{\lambda}}$. Here, its components are orthonormal, i.e., $(\phi_{\lambda,m})^{\dagger}\phi_{\lambda,n}=\delta_{mn}$, where $m,n=1,\ldots, N_{\l}$ labels the degenerated eigenvectors and hence $\Phi^{\dagger} \Phi=1$. The (nonabelian) Berry connection is defined as ${\bm A}_{\lambda,mn}=(- \i \phi_{\lambda,m} ^{\dagger}\bm{\nabla}_\vk \phi_{\lambda,n})/N_{\lambda}$~\cite{Wilczek:1984dh}. The row vector $\Phi'_{\lambda}$  and Berry connection $\bm{A}'_\lambda$ associated with $\phi'_{\lambda}$s can be defined similarly. A different choice in eigenfunctions corresponds to the transformation
\begin{align}
\label{phi-gauge}
&\Phi_{\l}\to \Phi_{\l} U^\dagger_{\l}\,, \quad \Phi'^\dagger_{\l} \to U'_{\l} \Phi^\dagger_{\l} \,, 
\end{align}
where $U_{\l}$ and $U'_{\l}$ are $U(N_{\l})$ matrices. The connection ${\bm A}_{\lambda}$ and ${\bm A}'_{\lambda}$ will transform as $\vA \to U \vA U^{\dagger} -\i U \bm{\nabla}_\vk U^\dagger$ and $\vA' \to U' \vA' U'^{\dagger} -\i U' \bm{\nabla}_\vk U'^\dagger$. Here and hereafter, the subscript $\l$ and indices $m,n$ are omitted when clear from the context. The Berry structure of a single-particle state is characterized by the flux of the trace of Berry magnetic field $B^{i}=\epsilon^{ijk}F_{jk}$ and $ F_{ij}=\partial_{i}A_{j}-\partial_{j}A_{i}+ \i[A_{i},A_{j}]$ on the FS:
\begin{align}
\label{eq:q-def}
q= \frac{1}{4\pi} \int_{\rm FS} {\rm d}\bm{S} \cdot \tr {\bm B} =\frac{1}{4\pi} \int_{\rm FS} {\rm d}\bm{S}\cdot\left({\bm \nabla}_{\vk}\times\tr{\bm A}\right)\,,
\end{align}
and similarly for $q'$. Here $2 q=0,\pm 1$ is related to the first Chern number. A nonzero value of $q$ indicates that $\Phi$ cannot be defined globally on the FS. Without degeneracy, $\vA$ and $\vA'$ reduce to the Abelian Berry connection. We define the generalized pairing monopole charge as $\Delta q\equiv q-q'$.

Next, we connect $\Delta q$ to the topological number of the nodes of the projected gap function, $\tilde{M}^{\dagger}_{mn}\equiv \phi^{\prime\dagger}_m M^\dagger \phi_{n}$ (at given $\lambda$). Recall that the topology of a superfluid is characterized by the circulation of superfluid velocity around defects in real space where the superfluid is absent. The real-space superfluid velocity is defined by the gauge-invariant combination of the phase gradient of the gap function and the gauge field. Analogously, we define the ``momentum-space" velocity field as
\begin{align}
\label{u-def}
\vu\equiv  \bm{\nabla}_{\vk} \alpha-\tr({\bm A}-{\bm A}')\, , 
\end{align}
where $\alpha=-\i(\log \det{\tilde M^{\dagger}})/N_{\lambda}$ is the phase of $\det{\tilde M^{\dagger}}$. It is easy to verify that ${\bm u}$ remains unchanged under the transformation~\eqref{phi-gauge}.

\begin{figure}[t]
\centering
\includegraphics[bb=0 0 780 420, width=8cm]{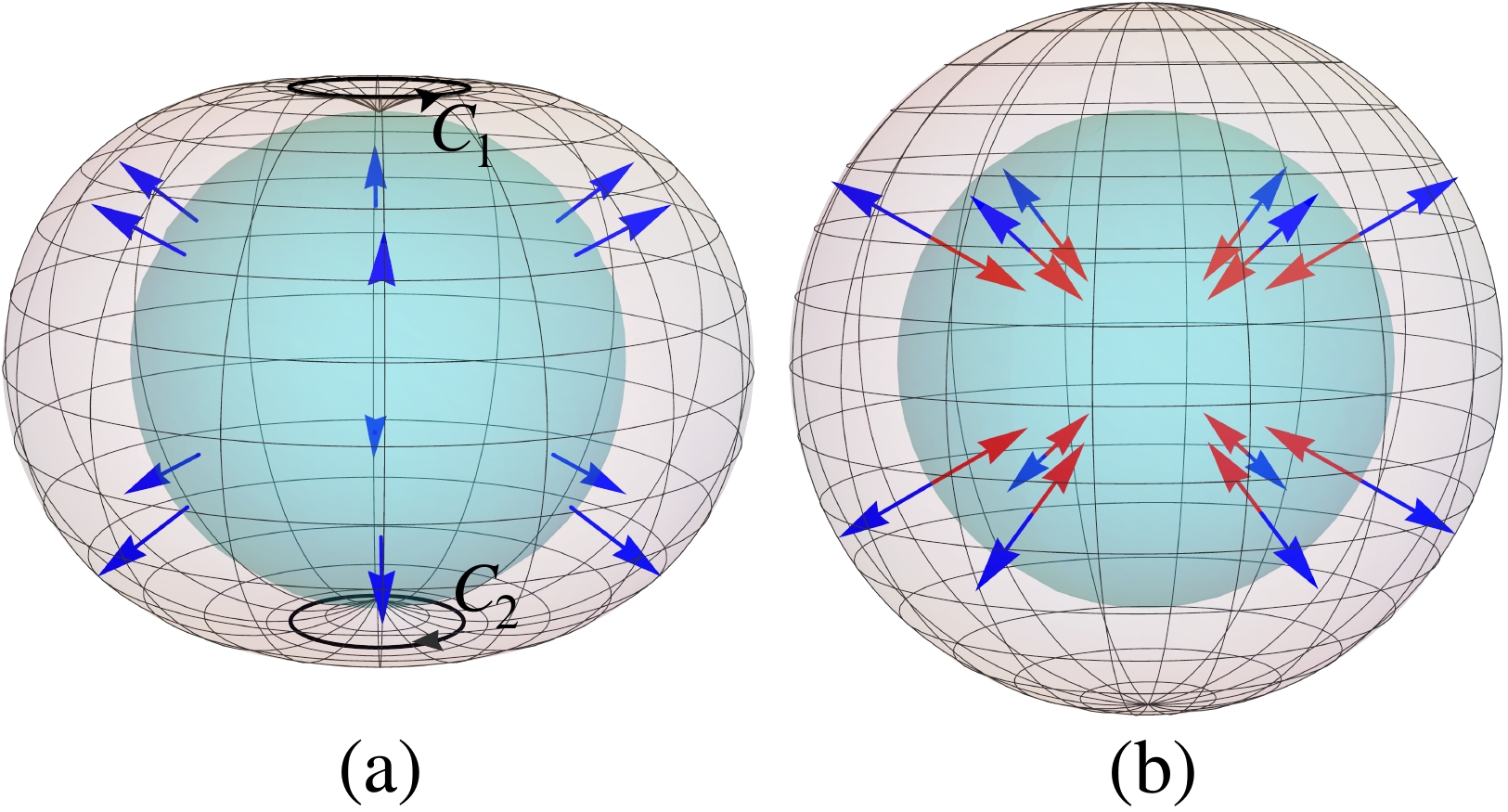}
\caption{
\label{fig:FS}
Schematic representation of the equal energy eigenvalue contour for $\sqrt{(k-\mu)^{2}+\lambda(\hk)\,\Delta^{2}_{0}}$ around the FS for the transverse phase (the opposite chirality pairing) of spin-one color superconductor.
The blue and red arrows illustrate the pairing of Berry flux from the chirality (spin) and color contributions, respectively. (a) The polar phase where the gap closes at the nodes at the north pole $\hat{\bm{k}}_{\rm node,1} = \hat{\bm{k}}_{z}$ and the south pole $\hat{\bm{k}}_{\rm node,2} = -\hat{\bm{k}}_{z}$, lacking color contribution to the pairing Berry flux. (b) CSL phase, which is fully gapped due to cancellation between color and chirality Berry fluxes. 
}
\end{figure}

If nodes exist at $\hat{\vk}=\hat{\vk}_{{\rm node}, N}$ $(N=1,2,\ldots)$ where the gap $\tilde{M}$ (and $\lambda$) vanishes and $\alpha$ is ill defined, we consider the circulation of $\vu$ along the infinitesimally small oriented loop (by the right-handed rule) $C_{N}$ around $\hat{\vk}_{{\rm node}, N}$ [see  Fig.~\ref{fig:FS}~(a)]. A node is termed ``topological" when the circulation is nonzero. Summing the circulations yields
\begin{align}
\label{g-def}
g&\equiv\frac{1}{2\pi}\sum_N \oint_{C_N} {\rm d}\bm{t} \cdot \vu
=\frac{-1}{2\pi} \iint_{\rm FS} {\rm d}\bm{S} \cdot (\bm{\nabla}_{\vk}\times {\bm u})\,,
\end{align}
where we have reversed the loop $C_{N}$ and applied the Stokes theorem. Away from the nodes, ${\bm \nabla}_{\bm{k}}\times {\bm \nabla}\alpha$ vanishes, and the ``vorticity" in momentum space ${\bm \nabla}_{\bm{k}}\times {\bm u}$ coincides with the trace of the pairing Berry flux $-{\bm \nabla}_{\bm{k}} \times ({\bm A}-{\bm A}')$. This leads to the key relation 
\begin{align}
\label{g-relation}
g&=2\Delta q=2(q-q')\,,
\end{align}
which generalizes the early relation~\eqref{eq:g-spin} and is new in literature. Crucially, $g$ is determined not by $\Delta q_{{\rm ch}}$ but by $\Delta q$, which incorporates contributions to the pairing Berry flux from additional quantum numbers. This generalization allows us to analyze the topological aspects of spin-one color superconductivity.

\sect{
Spin-one color superconductor}
We begin by examining the pairing between quarks with opposite chirality (transverse phase) in light of relation~\eqref{g-relation}. Focus on pairing between the right-handed quark $\psi_{\rm R}$ and left-handed quark $\psi_{\rm c L}$, which decouples from the one between $\psi_{\rm L}$ and $\psi_{\rm c R}$ (LR sector)~\cite{Schafer:2000tw,Schmitt:2004et}, we derive the reduced Hamiltonian of the form \eqref{eq:gen-ham}. Here, ${\cal H}_{0}= {\bm \sigma}\cdot \vk-\mu$ and ${\cal H}_{0{\rm c}}^{\prime}=-{\cal H}_{0}(-\vk)$, where $\mu$ is chemical potential and the gap matrix $M$ reads
\begin{align}
\label{M-CSC}
M =(P_{+}\,\si^{\perp}_{i})\, \Delta_{ia} J_a \,,
\end{align}
where $(J_{a})_{bc} =- \i \epsilon_{abc}$ are the antisymmetric color matrices with $a$, $b$, and $c$ denoting color indices. The spin matrix $\sigma^{\perp}_{i}=(\delta_{ij} - \hat{k}_i\hat{k}_j)\sigma_{j}$ is transverse to $\hat{\bm{k}}$ (hence the name ``transverse" phase). At this point, we consider a general $3\times3$ matrix $\Delta_{ia}$. Note that $M$ introduced by Eq.~\eqref{M-CSC} satisfies Eq.~\eqref{M-commute}, and the relation~\eqref{g-relation} is applicable. This also implies that the eigenfunctions of $M M^\dagger$ and $M^\dagger M$ can possess definite helicity. With $P_{+}= (1+\bm{\sigma} \cdot \hat \vk)/2$ in $M$, we find $M\xi_{\rm R}=0$ and $M^{\dagger}\xi_{\rm L}=0$ for normalized right- and left-handed spinors, $\xi_{\rm R}$ and $\xi_{\rm L}$. Noting $M$ is proportional to the director product of color and spin matrices, we write the modes of interest as
\begin{align}
\label{eigenfunction-color-spin}
\phi_{\rm R} (\hk) =c(\hk) \otimes \xi_{\rm R}(\hk)\,, \quad \phi^\prime_{\rm L} (\hk) =c^\prime(\hk) \otimes \xi_{\rm L}(\hk)\,,
\end{align}
where $c$ and $c'$ are three-vectors in the color space, normalized as ${\bm c}^{*}\cdot {\bm c}=1$.

To determine $c$ and $c'$, we substitute Eqs.~\eqref{M-CSC} and \eqref{eigenfunction-color-spin} into Eq.~\eqref{eq:eigen_problem_MMd}. Note
\begin{align}
\label{l-def}
 (P_{+}\bm{\sigma}_\perp)\xi_{\rm L} ={\bm l}_{-}\xi_{\rm R} \,,
 \quad
  (P_{+}\bm{\sigma}_\perp)^\dagger \xi_{\rm R} = {\bm l}_+\xi_{\rm L}\, ,
\end{align}
where the explicit expression for ${\bm l}_{\pm}$ depends on the choice of $\xi_{\rm R,L}$ [see Eq.~\eqref{l-gauge} below]. Interestingly, ${\bm l}_{\pm}$ coincides with the eigenstate of the photon helicity operator ${\bm S}\cdot \hk$ with the eigenvalue $\pm 1$, where $(S_{i})_{jk}=-\i\epsilon_{ijk}$ is the generator of ${\rm SO}(3)$ spatial rotation \cite{bialynicki1996v,Yamamoto:2017uul}. For completeness, we define $\bm{l}_{0}=\hk$ as the helicity zero state.

Now, Eq.~\eqref{eq:eigen_problem_MMd} reduces to the eigenvalue problem,
\begin{align}
\label{NN-c}
NN^{\dagger} c=\l c\,,\quad N^{\dagger}N c=\l c'\,,
\end{align} 
where the color matrix $N$ is given by
\begin{align}
\label{n-def}
N\equiv \xi^{\dagger}_{\rm R}M\xi_{\rm L}={\bm n}\cdot{\bm J}\,,
\quad n_{a}= (l_-)_{i}\Delta_{ia} \,.     
\end{align}
Using the definition of ${\bm J}$, we explicitly have
\begin{align}
\label{NNd}
(N N^{\dagger})_{ab}= ({\bm n}\cdot {\bm n}^{*}) \delta_{ab}- n^{*}_{a}n_{b}\,,
\end{align}
where $n_{a}=\delta_{ai}n_{i}$ is a three-vector in color space. Equation~\eqref{NNd} implies that if $c_{1}$ and $c_{2}$ are (1) orthonormal $c^{*}_{1} \cdot c_{2}=0$ and (2) orthogonal to $ n^{*}_{a}$( $c^{*}_{1}\cdot n^{*}=c^{*}_{2}\cdot n^{*}=0$), then $c_{1}$ and $c_{2}$ are degenerate nonzero modes of $NN^{\dagger}$ with eigenvalue $\lambda = {\bm n}\cdot{\bm n}^{*}$.

The nonabelian Berry connection of the modes~\eqref{eigenfunction-color-spin} now becomes $\bm{A}_{mn} =  \left( \bm{A}_{\co,mn} + \delta_{mn}\bm{A}_{\rm R} \right)/N_\lambda$ (similar for ${\bm A}'$), where $\bm{A}_{\co, mn} = -\i c^{\dagger}_{m} \bm{\nabla}_{\vk} c_{n}$ represents the color contribution to the Berry gauge field. Accordingly, $\Delta q$ decomposes as $\Delta q = \Delta q_{\rm ch} + \Delta q_{\rm co}$, where $\Delta q_{\rm co} \equiv q_{\rm co} - q_{\rm co}'$. For later convenience, the projected gap matrix is recorded as
$\tilde{M}_{mn}=  c'_{m,a}\,N_{ab}\,c_{n,b}$.

Once the order parameter $\Delta_{ia}$ is specified, we can evaluate the circulation of velocity and Berry flux according to Eq.~\eqref{g-relation} explicitly. The relevant expressions for different phases are summarized in Table~\ref{table:summary}.

\begin{table}[t]
\centering
\begin{tabular}{|c|c|c|c|}
\hline
~
& Polar
& {\it A} phase 
& CSL \\
\hline
$\Delta_{ia}$
& $\delta_{i3}\delta_{a3}$ 
&$(\delta_{i1}+\i \delta_{i2})\delta_{a3}$
&$\delta_{ia} $\\
\hline
$\bm{n}$ 
& $l_{-,3}\bm{\he}_{3}$ 
&$(l_{-,1}+\i l_{-,2})\bm{\he}_{3}$
&${\bm l}_{-} $\\
\hline
\hline
$\Delta q_{\rm ch}$
&
$1$
&
$1$
&
$1$
\\
\hline
$\Delta q_{\co}$
&
$0$
&
$0$
&
$-1$
\\

\hline
$g$
&
$2$
&
$2$
&
$0$
\\
\hline
$\Delta q_{0,\co}$
&
$0$
&
$0$
&
$2$
\\
\hline
\end{tabular}
\caption{
\label{table:summary}
The order parameter ($3\times3$ matrix) $\Delta_{ia}$ for different transverse phases of the spin-one color superconductor. 
With the expression for vector $\bm{n}$ given by Eq.~\eqref{n-def}  for the opposite chirality (third row),
we compute nonzero eigenvalue $\l={\bm n}\cdot {\bm n}^{*}$ and eigenvectors of Eq.~\eqref{NN-c}.
}
\end{table}

\sect{Topological nodal structure of color superconductor}
We first discuss the polar phase (see Table~\ref{table:summary}). 
The quasiparticle excitation $E=\sqrt{(k-\mu)^{2}+\Delta^{2}_{0}\lambda}$ is illustrated in Fig.~\ref{fig:FS} (a), where $\lambda=\sin^{2}\theta$ (see also Refs.~\cite{Schafer:2000tw,Schmitt:2004et}). The nodes are at the north ($\theta=0$) and south poles ($\theta=\pi$). We find $c_{1}=(1,0,0)$ and $c_{2}=(0,0,1)$, indicating $\Delta q$ is solely given by the chirality contribution: $\Delta q=\Delta q_{{\rm ch}}=1$.

The nonzero total velocity circulation implies obstruction in defining the phase of the gap $\tilde{M}$ globally due to the Berry structure of the pairing. 
To demonstrate this, we choose the helicity basis that behaves regularly on the FS. Near the north pole, we use $\xi_{\rm R} = (\cos(\theta/2)\,, \e^{\i \phi} \sin(\theta/2))$ and $\xi_{\rm L} = (\e^{-\i \phi}\sin(\theta/2)\,, -\cos(\theta/2))$, and resulting Berry gauge fields are $\vA_{\rm R}=-\vA_{\rm L}=(1-\cos\theta)/(2 k \sin\theta)\he_{\phi}$. Evaluating Eq.~\eqref{l-def} gives
\begin{align}
\label{l-gauge}
\bm{l}_{\pm}= 
\left( \cos\theta \cos\phi \pm\i \sin\phi\,,\ \cos \theta \sin\phi \mp\i \cos\phi\,,\ \sin\theta \right)\,.
\end{align}
Near the south pole, we switch to another gauge $\xi_{\rm R}\rightarrow \e^{-\i \phi}\xi_{\rm R}$ and $\xi_{\rm L}\rightarrow \e^{\i \phi}\xi_{\rm L}$. Thus, $\vA_{\rm R}=-\vA_{\rm L}=(-1-\cos\theta)/(2 k \sin\theta)\he_{\phi}$ and ${\bm l}_\pm \to \e^{\mp 2\i\phi}{\bm l}_{\pm}$. We find $\det (\tilde{M}^{\dagger})=l_{+,3}$, leading to $\a=\phi$ and $\a=-\phi$ near the north and south poles, respectively. The circulation around each node, obtained by integrating over $\bm{\nabla}_{\bm k}\alpha$, is $1$ for both. The sum of them, arising from the ``gauge difference" in $\alpha$ around the nodes, is $g=2=2\Delta q$, consistent with Eq.~\eqref{g-relation}.

Turning to the {\it A} phase (see Table~\ref{table:summary}), we also have $\Delta q=\Delta q_{{\rm ch}}=1$. Explicit calculations reveal $\l=2(1+\cos\theta)$ [for LR sector, $\l=2(1-\cos\theta)$], with a single node at $\theta=\pi$. With the basis that is regular near the south pole, we find ${\rm det}{\tilde M}\propto \e^{2\i\phi}(1+\cos\theta)$, indicating $g=2$. Despite the difference in the number of nodes and circulation, the relation~\eqref{g-relation} holds for both the polar and {\it A} phases.

The order parameter of the single-flavor color superconductor is similar to that of the superfluid ${\rm ^3He}$, where the rotation in the orbital angular momentum parallels rotation in the color space. The {\it A} phase of the superfluid \He exhibits two topological nodes at the north and south poles, with opposite circulation $1$ and $-1$, resulting in a different topological number $g=0$~\cite{vollhardt2013superfluid,Schnyder_2015,PismaZhETF.43.428,Bevan_1997,Volovik:1998wx}.

The Berry structure of the CSL phase contrasts sharply with that of the polar and {\it A} phases. We verify that $c_{1,a}\propto \delta_{ai}l_{-,i}=l_{-,a}$ and $c_{2,a}\propto l_{0,a}$ are degenerate eigenmodes of $NN^{\dagger}$ given by Eq.~\eqref{NNd} in $\lambda=2$ using ${\bm l}_{-} \cdot {\bm l}_{-}={\bm l}_{-}\cdot{\bm l}_{0}=0$. Since $l_{-,i}$ and $l_{0,i}$ are eigenvectors of the photon helicity operator, $c_{1,a},c_{2,a}$ are eigenvectors of ``color helicity" operator ${\bm J}\cdot \hk$ with color helicity $-1$ and $0$, which coincide with their color monopole charge. Therefore, $q_{\co}=(-1)/N_{\lambda}=-1/2$ for $N_{\lambda}=2$ and similarly $q'_{\co}=1/2$. The color contribution is $\Delta q_{\co}=-1$ and it precisely cancels that from the chirality $\Delta q_{\rm ch}=1$  [see Fig.~\ref{fig:FS} (b)], implying $g=0$, which explains the puzzling fact that the CSL phase is fully gapped. Similar to the CSL phase, one can examine the planar phase $\Delta_{ia}=\delta_{i1}\delta_{a1}+\delta_{i2}\delta_{a2}$, which is fully gapped with $g =0$.

\sect{Gapless excitation}
We now demonstrate that the presence of $\Delta q_{{\rm ch}}$ imposes a constraint on the color monopole charge for the gapless excitation of the pairing state. Those excitations can be expressed in terms of the zero eigenvectors of $NN^{\dagger}$ and $N^{\dagger}N$, denoted by $c_{0}$ and $c'_{0}$, as $\phi_{0,{\rm R}}=\le(c_{0}\otimes\xi_{\rm R},0\ri)^{\rm T}$ and $\phi'_{0,{\rm L}}=\le(0,c'_{0}\otimes\xi_{\rm L}\ri)^{\rm T}$. The Berry connection for the gapless mode is given by ${\bm A}_{0}={\bm A}_{0,\co}+{\bm A}_{\rm R}$ (similar for ${\bm A}'$), where $\bm{A}_{0,\co} = -\i c^{\dagger}_{0} \bm{\nabla}_{\vk} c_{0}$ is the color contribution.

Since $c_{0,1,2}$ forms a complete basis in the color space, we can easily show (see Appendix) $ N_{\l} q_{\co}+q_{0,\co}=0$. Therefore, $\Delta q_{\co}=-\Delta q_{0,\co}/2$ ($N_{\l}=2$) and $\Delta q=\Delta q_{{\rm ch}}-\Delta q_{0,\co}/2$. Rewriting Eq.~\eqref{g-relation} gives the relation
\begin{align}
\label{q-zero}
2\Delta q_{{\rm ch}}= g+\Delta q_{0,\co}\,. 
\end{align}

The sum rule~\eqref{q-zero} reveals the rich physical consequence of nonzero $\Delta q_{{\rm ch}}$ for the pairing fermions with additional quantum numbers.

(1) Scenario A: $2\Delta q_{{\rm ch}}$ is saturated by the total circulation $g$ with $q_{0,\co}=0$\,. Under this scenario, the system cannot be fully gapped. This applies to the transverse polar and {\it A} phase. We have checked $c_{0}=c'_{0}=(0,0,1)$ in those cases, corresponding to the blue quark not participating in the pairing. 

(2) Scenario B: $2\Delta q_{\rm ch}=\Delta q_{0,\co}$ and $g=0$. This is exemplified by the CSL phase, where the system is fully gapped ($g=0$). In this case, $c_{0,a}\propto l_{+,a}$ and $c'_{0,a}\propto l_{-,a} $, hence $q_{0,\co}=1$ and $q'_{0,\co}=-1$. Therefore, $\Delta q_{0,\co}=2$ as expected. The gaples excitations $\phi_{0,{\rm R}}$ and $\phi'_{0,{\rm L}}$ have unusual Berry monopole charges $3/2$ and $-3/2$, as they carry both color helicity ($\pm 1$) and spin helicity ($\pm 1/2$).

While the analog of scenario A was known in the context of the superconductor phase of doped Weyl metals~\cite{Li-Haldane-PRL18}, scenario B has not been recognized to date.

\sect{Longitudinal vs transverse phase}
Let us perform a parallel analysis for the longitudinal phase, to say the pairing between right-handed quarks. Here, in Eq.~\eqref{eq:gen-ham}, we use ${\cal H}_{0}= {\bm \sigma}\cdot \vk-\mu$ and ${\cal H}_{0{\rm c}}^{\prime}={\cal H}_{0}(-\vk)$ and $M=P_+\,(\hat{k}_{i}\Delta_{ia}J_{a})$. The relevant eigenmodes can be written $\tphi =\tc \otimes \xi_{\rm R}$ and $\tphi' = \tc' \otimes \xi_{\rm R} $ [cf., Eq.~\eqref{eigenfunction-color-spin}]. Here, $\tc$ is the eigenvector of the color matrix $\tN\tN^{\dagger}$, where $\tN$ is obtained by replacing ${\bm l}_{-}$ in Eq.~\eqref{n-def} by ${\bm l}_{0}$ [note, $(P_{+}\hk) \xi_{\rm R}={\bm l}_{0}\xi_{\rm R}$]. Obviously, $q_{{\rm ch}}=q'_{{\rm ch}}=1/2$ and $\Delta q_{{\rm ch}}=0$.

For {\it A} phase, the color vectors $\tc_{1,2}$ are constant and $q_{\co}=q'_{\co}=0$, yielding $\Delta q=0$. The eigenvalue $\l=\sin^{2}\theta$ has nodes at the north and south poles, where the circulations are $1$ and $-1$, respectively (the same as the {\it A} phase of $^3{\rm He}$). Thus, $g=0$, in accordance with the relation~\eqref{g-relation}. For the CSL phase, we confirmed that it is fully gapped ($\lambda=\hk^{2}=1$) with $\tc_{1,a}=l_{+,a}$ and $\tc_{2,a}=l_{-,a}$, which carry opposite color helicity, $1$ and $-1$, respectively, resulting in $q_{\co}=0$. We further verify Eq.~\eqref{q-zero} by noting $\tc_{0}=l_{0}$ that carries zero color Berry flux. The gapless excitation has Berry monopole charge $\pm 1/2$ instead of $\pm 3/2$ that we found in the transverse CSL phase.

\sect{Summary and outlook}
We have explored the impact of Berry flux on the BCS state formed by fermions carrying nonzero Berry monopole charge, particularly quarks in QCD at high density. Previous analyses on the topological properties of color superconductors are mostly concentrated on the CFL phase (see Refs.~\cite{Nishida:2010wr,Alford:2018mqj,Chatterjee:2018nxe,Cherman:2018jir,Hirono:2018fjr,Hidaka:2022blq,Hayashi:2023sas,Hayata:2024nrl}). The rich Berry-topological structure of spin-one single-flavor pairing states has remained largely unexplored. Furthermore, we demonstrate that, although the transverse and longitudinal color superconductive phases at a given order parameter share the same global symmetry-breaking pattern, they can be distinguished by the difference in the topological structure of nodes and the Berry monopole charges of quasiparticles.

It would be interesting to explore further physical consequences of the gapless modes with unusual Berry monopole charge $\pm 3/2$ in the CSL phase (e.g., the response to external magnetic and electric fields), which exhibits electromagnetic superconductivity \cite{Schmitt:2003xq,Feng:2009vt}.
Given the close connection between Berry monopole charge and quantum anomaly, it is worthwhile to investigate the implications of our findings on the criteria for anomaly matching in dense quark matter.

In this study, we neglect the effects of quark mass on the Berry curvature, which is a valid approximation for $u,d$ quarks. We anticipate that our analysis captures the qualitative features of the Berry structure even in the three-flavor case, provided the strange quark mass $m_s$ is sufficiently large to induce a significant Fermi-momentum mismatch--thereby suppressing cross-flavor pairings--but still small enough to justify the ultrarelativistic approximation. Investigating the impact of $m_{s}$ will be reserved for the future.

Fermionic quasiparticles with internal degrees of freedom analogous to color emerge in a variety of condensed matter systems, offering a broader context for the phenomena studied in this Letter. For instance, the gas of ultracold atoms with multiple hyperfine states, such as alkaline-earth Fermi gases, exhibit emergent ${\rm SU}(N)$ $(N>2)$ symmetry~\cite{Cazalilla:2014wfa} that leads to a rich structure of paired states, including the formation of ``color superfluidity" analogous to dense QCD matter~\cite{PhysRevA.74.033604,PhysRevLett.98.160405}. Other systems that may host fermionic excitations include the twisted bilayer graphene~\cite{PhysRevLett.128.227601} and multi-Weyl semimetals~\cite{Dantas:2019rgp}. These examples demonstrate the potential to extend the relevance of our findings to a wide range of physical systems.

\sect{Acknowledgments}We thank Dimitri Kharzeev, Qing-Dong Jiang, Zhen Liu, Andreas Schmitt, Misha Stephanov, Xin-yang Wang, Naoki Yamamoto, and Ho-Ung Yee for the helpful discussion. We thank Yukio Tanaka and Grigory Volovik for the variable comments. This work is supported by the U.S. Department of Energy, Office of Science, Office of Nuclear Physics Award No. DE-FG0201ER41195 (N.~S.). Y.~Y. acknowledges the support from NSFC under Grant No.12175282 and by CUHK-Shenzhen University Development Fund under the Grant No. UDF01003791. 

\sect{Data availability}No data were created or analyzed in this study.

\appendix
\section{Appendix: 
Proof of $N_\lambda q_{\co}+q_{0,\co}=0$}
\label{appendix}

Using $\bm{A}_{\co, mn} = -\i c^{\dagger}_{m} \bm{\nabla}_{\vk} c_{n}$, we compute the nonabelian Berry curvature [$\partial_i\equiv (\bm{\nabla}_{\vk})_i$ and omitting the color index]:
\begin{align}
(F_{{\rm co},{mn}})_{ij} 
&= \partial_i (A_{\co,mn})_{j} - \partial_j (A_{\co,mn})_{i} +\i \left( [(A_{\co})_i,(A_{\co})_j] \right)_{mn}  \notag\\
&=-\i \partial_i (c^\dagger_m \partial_j c_n) + \i \partial_j (c^\dagger_m \partial_i c_n)\notag\\
&\quad -\i \sum_{l=1,2} \left[ (c^\dagger_m \partial_i c_l) (c^\dagger_l \partial_j c_n) - (c^\dagger_m \partial_j c_l) (c^\dagger_l \partial_i c_n) \right] \notag\\
&=-\i (\partial_i c^\dagger_m \partial_j c_n) + \i (\partial_j c^\dagger_m \partial_i c_n) \notag\\
&\quad +\i \sum_{l=1,2} \left[ (\partial_i  c^\dagger_m c_l) (c^\dagger_l \partial_j c_n) - (\partial_j c^\dagger_m c_l) (c^\dagger_l \partial_i c_n) \right]\notag\\
&=-\i (\partial_i c^\dagger_m \partial_j c_n) + \i (\partial_j c^\dagger_m \partial_i c_n)\notag\\
&\quad +\i \sum_{\bar{l}=0,1,2} \left[ (\partial_i  c^\dagger_m c_{\bar{l}}) (c^\dagger_{\bar{l}} \partial_j c_n) - (\partial_j c^\dagger_m c_{\bar{l}}) (c^\dagger_{\bar{l}} \partial_i c_n) \right]\notag\\
&\quad -\i \left[ (\partial_i  c^\dagger_m c_0) (c^\dagger_0 \partial_j c_n) - (\partial_j c^\dagger_m c_0) (c^\dagger_0 \partial_i c_n) \right]\notag\\
&= - \i \left[ (\partial_i  c^\dagger_m c_0) (c^\dagger_0 \partial_j c_n) - (\partial_j c^\dagger_m c_0) (c^\dagger_0 \partial_i c_n) \right]\,,
\end{align}
where we have used $c_m^\dagger {\bm \nabla}_{\bm k} c_n = -({\bm \nabla}_{\bm k} c_m^\dagger) c_n $ %
and $\sum_{\bar{l}=0,1,2} c_{\bar{l},a} c^{*}_{\bar{l},b}=\delta_{ab}$. 
With similar algebra, we compute the trace as 
\begin{align}
{\rm tr}\,(F_{\rm co})_{ij} &= - \sum_{n=1,2}\i \left[ (\partial_i c^\dagger_n c_0) (c^\dagger_0 \partial_j c_n) - (\partial_j c^\dagger_n c_0) (c^\dagger_0 \partial_i c_n) \right] \notag\\
&= - \sum_{\bar{n}=0,1,2}\i \left[ (\partial_i c^\dagger_{\bar{n}} c_0) (c^\dagger_0 \partial_j c_{\bar{n}}) - (\partial_j c^\dagger_{\bar{n}} c_0) (c^\dagger_0 \partial_i c_{\bar{n}} ) \right] \notag\\
&\quad  + \i \left[ (\partial_i c^\dagger_0 c_0) (c^\dagger_0 \partial_j c_0) - (\partial_j c^\dagger_0 c_0) (c^\dagger_0 \partial_i c_0) \right] \notag\\
&= - \sum_{\bar{n}=0,1,2}\i \left[ (c^\dagger_{\bar{n}} \partial_i c_0) (\partial_j c^\dagger_0 c_{\bar{n}}) - (c^\dagger_{\bar{n}} \partial_j  c_0) (\partial_i c^\dagger_0 c_{\bar{n}} ) \right] \notag\\
&= \sum_{\bar{n}=0,1,2}\i \left[(\partial_i c^\dagger_0 c_{\bar{n}} ) 
 (c^\dagger_{\bar{n}} \partial_j  c_0)  - (\partial_j c^\dagger_0 c_{\bar{n}}) (c^\dagger_{\bar{n}} \partial_i c_0) \right] \notag\\
&= \i \left[(\partial_i c^\dagger_0 \partial_j  c_0) - (\partial_j c^\dagger_0 \partial_i c_0) \right] \notag\\
&= - F_{0, {\rm co}} \,.
\end{align}
Recalling that $q_{\rm co}$ is an average over degenerated states
, we obtain
\begin{align}
N_\lambda q_{\rm co} = - q_{0,{\rm co}}\,,
\end{align}
which was what we wanted.



\bibliography{ref}
\end{document}